\documentclass{article}
\usepackage[utf8]{inputenc}
\usepackage{amsmath}
\usepackage{graphicx}
\usepackage{geometry} 
\usepackage[
    backend=biber,
    style=numeric,
  ]{biblatex}
\usepackage{lineno}

\title{Social Learning and the Exploration-Exploitation Tradeoff} 
\author{Brian Mintz, Feng Fu\\
Department of Mathematics, Dartmouth College, Hanover, NH 03755, USA} 
\date{November 2022}

\begin{document}
\maketitle

\begin{abstract}
    Cultures around the world show varying levels of conservatism. While maintaining traditional ideas prevents wrong ones from being embraced, it also slows or prevents adaptation to new times. Without exploration there can be no improvement, but often this effort is wasted as it fails to produce better results, making it better to exploit the best known option. This tension is known as the exploration/exploitation issue, and it occurs at the individual and group levels, whenever decisions are made. As such, it is has been investigated across many disciplines. In this work, we investigate the balance between exploration and exploitation in changing environments by thinking of exploration as mutation in a trait space with a varying fitness function. Specifically, we study how exploration rates evolves by applying adaptive dynamics to the replicator-mutator equation, under two types of fitness functions. For the first, payoffs are accrued from playing a two-player, two-action symmetric game, we consider representatives of all games in this class and find exploration rates often evolve downwards, but can also undergo neutral selection as well. Second, we study time dependent fitness with a function having a single oscillating peak. By increasing the period, we see a jump in the optimal exploration rate, which then decreases towards zero. These results establish several possible evolutionary scenarios for exploration rates, providing insight into many applications, including why we can see such diversity in rates of cultural change. 
    
    \vspace{0.5cm}
    
    \textbf{Keywords:} Adaptive dynamics, exploration/exploitation, Social learning, evolutionary dynamics, mutation rate evolution. 
\end{abstract}

\section{Introduction}
In any learning process, individuals leverage past information along with the opinions of others to decide their best action. Broadly speaking, one can either continue using a strategy that has worked, or try a new approach. While exploration is necessary to discover better strategies, it often results in wasted effort, so it is usually better to exploit the best known strategy. These opposing approaches are very general, applying any time a decision must be made. As such, this concept is relevant across scales, both at the individual such as animal or cells, and group levels, in a wide range of areas from biology to economics \cite{berger2014exploration}. Much work has gone into studying this issue from a variety of perspectives. 

One can think of mutation as exploration in the space of genomes. Since all lifeforms replicate their genetic information, the study of mutation rates has been a longstanding area in biology with significant implications \cite{lynch2016genetic}. For example, studies have investigated the mechanisms for viral RNA repair and phenotypic switching in bacteria \cite{domingo2021mutation, liberman2011evolution}. There has also been considerable theoretical work on mutation rates in sexually reproducing organisms, finding higher mutation rates can be selected for or against depending on the situation \cite{burger1999evolution, romero2019elevated}. Beyond the level of individual cells, decision making in humans has been studied in the exploration/exploitation framework, including its neuroscientific underpinnings \cite{navarro2016learning, laureiro2010neuroscientific}. Additionally, this approach has been employed in several areas of ecology, including foraging and analyzing host-parasite or predator-prey systems, \cite{eliassen2007exploration, mgonigle2009mutating, monk2018ecology}.


Computer scientists have also investigated the balance between exploration and exploitation through evolutionary algorithms, which feature a mutation parameter \cite{vcrepinvsek2013exploration, eiben1998evolutionary}. For example, particle swarm optimization is a technique that uses a collection of agents to discover optimal values in a complex space \cite{kumar2023balancing}. One approach, known as simulated annealing, decreases the exploration rate over time to concentrate the population around the global optimum. Yet another technique called reinforcement learning has individuals track the performance of a set of possible actions over time to determine the optimal choice \cite{yen2002coordination}. In this framework, one makes an explicit policy for whether new actions are chosen to update these values, exploration, or the current best value is used, exploitation. This area has seen increasing interest from its application to artificial intelligence. 
 
Lastly, there is a significant history of studying exploration/exploitation in economics \cite{almahendra2015exploration}. Applications include theories of firm's flexibility and understanding product development and innovation, \cite{mathias2018managing, greve2007exploration, gilsing2006exploration}. Additionally, March's seminal model of mutual learning in organization, where an individuals and the firm learn from each other dynamically, has been extensively studied and generalized over the last few decades in management science, \cite{march1991exploration, bocanet2012balancing, lazer2007network, posen2012chasing}.

In this project, we ask how evolution would solve the exploration/exploitation issue, by allowing natural selection to operate on an exploration parameter in a variety of contexts. Specifically, we investigate the evolutionary forces on exploration rates in dynamic environments, as in \cite{nilsson2002optimal} and \cite{ben1993relationship}. We consider two different ways of modeling environmental change. The first uses a feedback mechanism between the strategies in a population and the fitness of those strategies, encoded as the average payoff of an individual when interacting with other players uniformly at random in a population playing a two-player two-action symmetric game. This approach is grounded in the tradition of evolutionary game theory. The other scenario we consider in this work is explicitly representing the fitness of each strategy by a time dependent function. In particular, we consider the fitness landscape with to have a single peak of some width, and whose location oscillates in time in some regular manner, similar to  \cite{ishii1989evolutionarily}. This case represents the fact that few environments are static in time, and often undergo periodic changes \cite{shu2022eco, wang2020eco}. For example, if we think of traits as preferred nesting sites in space, then the changing fitness could apply to the study of migration or dispersal.


\section{Methods}
We think of the set of actions an individual could take as a bounded continuous set, specifically real numbers in the unit interval, and the best action as a trait. This may seem restrictive, but up to linear transformations it can capture any bounded trait one can reasonably assign a number to, for example an organism's height or weight. By putting traits in a space, we can ensure exploration is local, with an exploration kernel to describe the probability distribution of an individual's trait in the near future given its current trait and some exploration rate $u$. In this work we consider a normal distribution with variation equal to $u$. Specifically, the model we will use for population dynamics is the replicator mutator equation: \begin{equation}\label{RME}
    \frac{d}{dt} \vec{x} = (Q(u)F(\vec{x},t)-\phi I) \vec{x}
\end{equation} where $\vec{x}$ is the trait distribution, $Q(u)$ gives the probability of exploration from one trait to another based on a exploration rate parameter $u$, $F(\vec{x},t)$ is a diagonal matrix with $ii$th entry the fitness of trait $i$ given the population distribution $\vec{x}$ and time $t$, and $\phi$ is the average fitness in the population (introduced as in the classical replicator equation to ensure the vector $\vec{x}$ sums to one). Essentially this equation makes individuals reproduce according to their fitness, for example by being imitated through social learning, and explore by the matrix $Q(u)$, depending on their exploration rate $u$. This framework is built on vectors, so the trait space is necessarily finite. Consequently, exploration cannot occur outside of the boundary, which we handle by truncating these values. Alternatively, one could accumulate them at the endpoints, or shift them to the other endpoint, making the trait space circular \cite{mintz2022point}. 

Equation (\ref{RME}) gives the short term population dynamics, and to represent the long term dynamics on exploration rate, we use the approach of adaptive dynamics. In this method, one considers the invasion fitness $f_x(y)$ of a mutant with trait $y$ in a population of individuals all with trait $x$, defined as their reproduction rate. It assumes mutations on the exploration rate is rare, so it suffices to consider monomorphic populations, as one trait will fixate before the next mutant arises. Further, if we assume these mutations are small, we can use a linear approximation $f_x(y) \approx f_x(x) + (y-x)\partial_y f_x(y)|_{y=x}$. If the term $\partial_y f_x(y)|_{y=x}$ is positive, $y-x$ must be as well for the trait to fixate, so the trait will evolve upwards. The same thing happens if this term is negative, so we can think of it as the rate of change of our trait. This framework was connected to the replicator-mutator equation in \cite{main_ref}, which determined $f_x(y)$ was the difference between the maximum eigenvalue of $Q(u)F(\tilde{x}, t^*)$ and the current average fitness, where $\tilde{x}$ is the stable distribution reached by the replicator-mutator equation under the exploration rate $x$, and $t^*$ is the time the mutant emerges. Using this, we can investigate the evolution of exploration rate if we specify the fitness function $F(\vec{x},t)$ and use the model of exploration specified above to define $Q(u)$. The code that implements this approach is available in the Github repository https://github.com/bmDart/exploration-rate-evolution.

To make fitness frequency-dependent, we consider a population playing a game. Each individual will then receive fitness that is the average payoff received over all possible interactions. Specifically, we will consider two-player, two-action, symmetric games, as these have a large degree of richness in their behavior. In these games, two players interact by each choosing one of two actions, $A$ or $B$, then receive a payoff dependent on the pair of actions chosen. There are four pairs, so one write can the payoffs to a player in the matrix \begin{center}
    \begin{tabular}{c|cc}
         & A & B  \\
         \hline
         A & $a$ & $b$ \\
         B & $c$ & $d$
    \end{tabular}
\end{center} where the rows correspond to a player's choice of $A$ or $B$, and the columns indicate the other player's choice of action. This class is called symmetric, as both players use the same matrix to determine their payoffs. It includes well known examples like the Prisoner's Dilemma (PD), if $b<d<a<c$, where individuals always do better by choosing the second action, even though the best outcome is both players choosing the first. Also included is the less intense version called the Hawk-Dove (HD), also called the Snowdrift, game, if $d<b<a<c$. In this game, the optimal action is the opposite of the other player's action, making this an anti-coordination game. Another game this framework encompasses is known as the Stag-Hunt (SH) game, where $b < d < c < a$. Here the optimal action is the same choice the other player makes, so this is a coordination game. 

\begin{figure}
\centering
\includegraphics[width=4.5in]{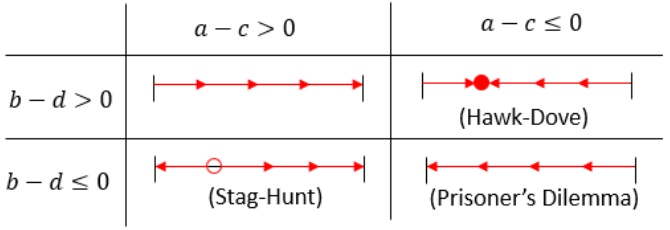}
\caption{The four possible cases for the evolution of strategies in two-player two-action symmetric games. Depending on the entries of the payoff matrix $\begin{bmatrix}
a & b\\  
c & d    
\end{bmatrix}$, either one endpoint will be attracting, or there will be an interior equilibrium that is either stable or unstable. The corresponding cases are labeled with their archetypal game. }
\label{AD-on-games}
\end{figure}

Strategies in these games can be complex, but if the game only consists of one round, and players have no information about each other, any strategy can be completely described by a probability distribution over the actions, a mixed strategy. Since there are only two possible actions, any strategy is a single number $x$, the probability of choosing the first action. Then the average payoff to a player with strategy $y$ interacting with a player of strategy $x$ is $$R(y,x) = ayx+by(1-x)+c(1-y)x+d(1-y)(1-x)$$ Since this is linear in $x$, the average payoff of an $y$ player interacting with a population with mean strategy $\bar{x}$ is just $R(y,\bar{x})$. Since this is also linear in $y$, so the average payoff over this population is $R(\bar{x},\bar{x})$. This function, $R(x,x)$, can give some insight, so we will refer to it as the population fitness of a strategy $x$. Interestingly, this can look differently within a class of games, for example $(a,b,c,d) = (2,0,2.5,1.5)$ and $(2,0,5,1.5)$ are two prisoner's dilemmas, yet the first makes $R(x,x)$ concave up while the other is concave down. We will see that increasing exploration rate often leads to a more spread out stable distribution, which moves the average strategy closer to 0.5, and see the effect this will have on a population's fitness. We can also apply Adaptive Dynamics, thinking of $R(y,x)$ as $f_x(y)$ the fitness of an invading strategy $y$ into a resident population of all $x$-players. Here we see the strategy should evolve according to $$\partial_y R(y,x)|_{y=x} = ax+b(1-x)-cx-d(1-x)$$ This is a line connecting $b-d$ at $x=0$ to $a-c$ at $x=1$, so there are essentially four cases depending on the relative signs of these terms, as shown in Fig.~\ref{AD-on-games} with representative games, and arrows indicating the dynamics of the invading strategies. Since the diagonal cases are essentially mirrors, we consider just the three games mentioned above. 

Lastly, we consider time-dependent fitness functions where the fitness landscape has a single peak, of some width, whose location oscillates at some frequency. In particular, we take the fitness at time $t$ to be the normal distribution with variance $0.1$ and mean $(1+\sin(\omega t))/2$, as this oscillates between the endpoints zero and one with period $\omega$. We investigate the effect of changing this period and also the variance of this distribution.

\section{Results}
\begin{figure}
\centering
\includegraphics[width=6.5in]{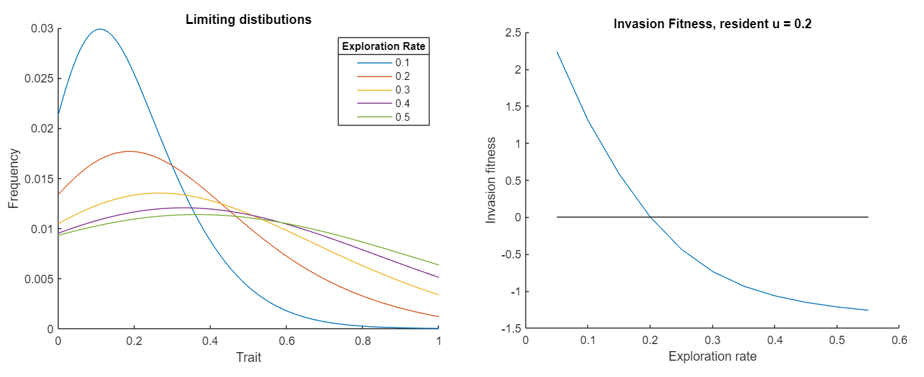}
\caption{On the left we see the stable distributions of the replicator mutator equation in the PD game for various exploration rates. The right plot shows the invasion fitness as a function of the invading exploration rate, for a resident value of 0.3. Since this is only positive to the left of the resident value, exploration rates can only evolve downwards. This is representative of all resident values. }
\label{PD}
\end{figure}

First, we used the payoff matrix $$\begin{bmatrix}
3 & 1\\
4 & 2
\end{bmatrix}$$ for the Prisoner's Dilemma, finding that the replicator-mutator equation stabilized at the distributions given in figure \ref{PD}. These show that lower exploration rates more closely exploit the optimal strategy of defection, as expected. However, those populations have lower fitness, as higher rates of choosing the second action, defection, are worse for the population overall, since the population fitness $R(x,x)$ is increasing for this game. Despite higher exploration rates leading to a population with greater fitness, we see the invasion fitness is only positive for lower exploration rates, so it can only evolve downward. This is a dilemma, as it is better to have a large exploration rate and flat trait distribution, but this will be selected against. 

\begin{figure}
\centering
\includegraphics[width=6.5in]{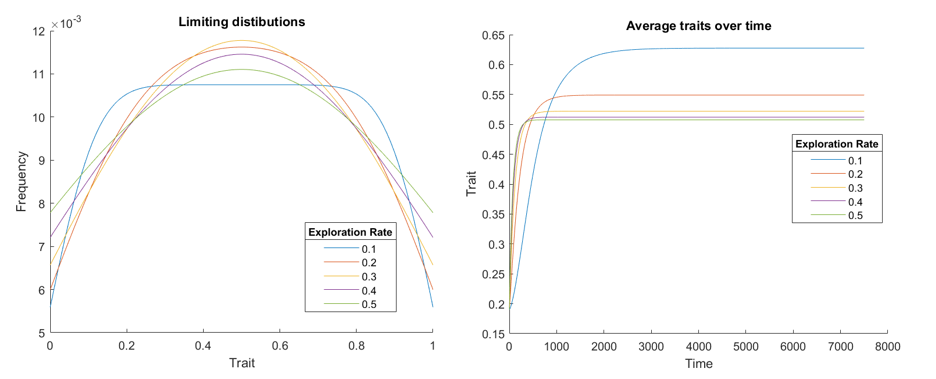}
\caption{In the Hawk-Dove Game, we see flatter stable distributions, the first plot, for large exploration rates for $c=0.5$. However, for $c \neq 0.5$, the average strategy does not reach the equalizer strategy $c$, shown in the second plot. }
\label{HD}
\end{figure}

The next game we considered is the Hawk-Dove game, with payoff matrix $$\begin{bmatrix}
1-c & 2\\
0 & 1
\end{bmatrix}$$ where $c$ is a parameter representing the cost of competing over a contested resource of value one. In this case, we see populations approach the equalizer strategy $cH+(1-c)D$, which makes all strategies have the same fitness, so there is no selection. Consequently, there is no selection on exploration rate, so it will be subject to neutral selection. This is consistent with the results of \cite{main_ref}, which found multiple mutation rates could coexist in this game. Like in the previous game, different stable distributions are reached for different exploration rates. Here we see increasingly uniform distributions as the exploration rate increases, in figure \ref{HD}, which is expected, as this represents larger exploration. Surprisingly, we see a dependence on the game parameter $c$. When this is not 0.5, the population does reach the equalizer strategy $c$, as seen by plotting the average strategy over time, also in figure \ref{HD}. This results in downward selection on exploration rate. Despite all Hawk-Dove games have the same strategy dynamics from the perspective of a single player, this population model demonstrates different effects depending on a parameter's value. Interestingly, despite exploration rates evolving downwards for $c \neq 0$, the population's fitness can either increase or decrease with exploration rate depending on if $c$ is above or below one half, as seen by considering the population fitness $R(x,x)$. Thus as in the PD, it is possible the exploration rate will evolve towards value that are worse for the population overall. 

The final game we considered was the Stag-Hunt, with payoff matrix $$\begin{bmatrix}
4 & 1\\
3 & 2
\end{bmatrix}$$ In this case, the population reaches unimodal distributions as in the Prisoner's dilemma, with individuals favoring one option more than the other. This is because the optimal action is to choose the same action as the other player, so the population becomes increasingly concentrated towards whichever pure strategy the initial mean was closer to. As such, the population will evolve away from the unstable equilibrium of 0.5, as in the single player dynamics. Depending on whether the initial mean is above or below 0.5, the population fitness is either increasing or decreasing with exploration rate, since this moves the mean strategy closer to a half, which is good if the population is concentrated around one but bad if it is concentrated around zero. Despite this, exploration rates can only evolve downwards in both case, so in this game too, exploration rates can evolve to less desirable levels. However in this case, selection becomes neutral for sufficiently large initial exploration rates, since the population becomes centered around 0.5. Thus, exploration rates that start large will drift up and down, but eventually become caught around zero.

The other type of fitness we considered in this work had fitness an explicit time-dependent function, with no dependence on the distribution of strategies in the population. Specifically, we took $f(x,t) = \exp(-(x-(1+\sin(\omega t))/2)^2)$ where $\omega$ is a parameter for the period of the oscillations. Here, one may also use the replicator-mutator equation to simulate the population dynamics, but now populations need not reach stable distributions. For example, a periodic fitness will lead to periodic changes in the population. Nonetheless, we can adapt the results of the model by considering a time averaged fitness. Since the fitness function does not depend on the frequency of strategies, an invading subpopulation with a novel exploration rate will grow independently of the resident population. Thus, the exploration rate leading to a higher average fitness will eventually fixate. Here one must use the geometric mean of fitness, as populations grow geometrically. This is because fitness is essentially a reproduction rate, which are multiplied, not added, together to aggregate over time periods, as is done in the geometric mean. Indeed, the order of the geometric and arithmetic mean might swap between two sets, for example $\{50,50\}$ and $\{100,1\}$. 

\begin{figure}
\centering
\includegraphics[width=4in]{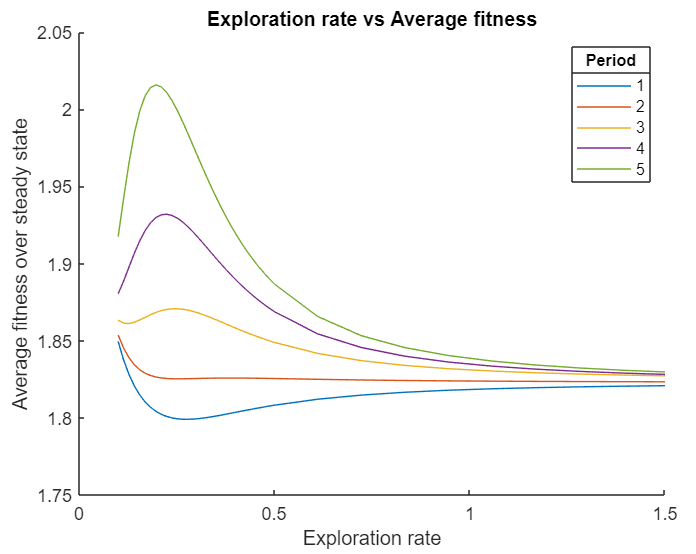}
\caption{For fitness that explicitly depends on time, we see both unstable and stable equilibria in the evolution of exploration rate. When environmental change is slow, corresponding to a long period, an intermediate level is optimal. Whereas for fast environmental change arbitrarily small exploration rates are optimal, though if the initial value is large enough, they will become arbitrarily large. }
\label{time-dep}
\end{figure}

In Figure \ref{time-dep}, we plot the time averaged fitness of each exploration rate, using fitness functions of various periods. For small periods, we see fitness is maximized around zero, decreasing with larger exploration rates until it reaches a local minimum then starts to increase. This means that for rapidly changing environments, it is best to have minimal exploration rate. However, if it starts above this minimum, exploration rates will increase arbitrarily high. This suggests some rates result in the population lagging behind the optimal strategy, to the extent that a uniform distribution is more effective. We see the opposite curve for sufficiently large periods, where environmental change is slow. Here, there is a local maximum at some nonzero exploration rate, indicating an intermediate level of exploration is optimal. Interestingly, as the period changes, the optimal exploration rate makes a jump from zero to an intermediate value. The exact period where this occurs and value the optimal rate jumps to would depend on specifics of the model, namely the type of curve defining the fitness. Further, optimal rate decreases with increasing period, that is, slower changing environments. This makes sense, as a sufficiently slow changing environment is effectively stable, for which arbitrarily small exploration rates are usually optimal. Comparable effects in the evolution of exploration rate are observed when the normal distribution has wider variance, indicating the generality of these results. Theoretically, one could also compute the time averaged fitness of the limiting exploration rates. When exploration rate is zero, the population will likely be entirely at the strategy that maximizes the time averaged fitness function, and when it is infinite, the strategy distribution will be uniform, so the time averaged fitness will simply be the average value of the function (which is constant in time, so equals its time average).

\section{Conclusion}
In this work, we investigated the evolution of exploration rate under variable selection, employing adaptive dynamics and the replicator-mutator equation. For frequency dependent fitness encoded by two-player/two-action, symmetric games, we found that exploration rates often evolve downward, but neutral selection is also possible. Despite this, it is possible in all games we considered for larger exploration rates to be more beneficial to the population. The precise form of the exploration rate's evolution can also depend on the game's parameters or the initial conditions, as in the HD and SH games respectively. This suggests that while the cases we studied are representative of the possible dynamics in this class of games, further richness could be observed in future study. However, we conjecture that this class of games is incapable of selecting for large exploration rates, as opposed to the more complicated class of two-player/three-action symmetric games, where it was found the Rock-Paper-Scissors game led to selection for an intermediate level of mutation \cite{rosenbloom2014frequency}. This is because cyclic dynamics in this game allow for a sub-population with multiple traits to remain resilient as the composition of the population changes. Since cyclic dynamics cannot be observed in the smaller class of games, it is likely that larger exploration rates cannot be selected for with these types of fitness function. Then with fitness function a single peak that oscillated according to some frequency, we found both attracting and repelling equilibria depending on the period. For fast changing environments, arbitrarily small exploration rates are optimal, though a sufficiently large initial exploration rate will evolve upwards. In contract, slow changing environments have intermediate optimal exploration rates, and evolution proceed towards this. As the rate of oscillation decreases further, this optimal value approaches zero \cite{yang2021dynamical}. 

This work could easily be extended to other trait topologies, by adapting the matrix $Q(u)$. For example, \cite{mintz2022point} explores how mutation rates can evolve upwards even in a fixed environment. This is found not just for traits in some interval, but also in a circular space or finite strings on a finite alphabet. Preliminary results showed the HD game with $c=0.5$ led to increasingly polarized distributions are exploration rates decreased, if exploration outside of the interval was accumulated at the endpoints. In addition, making the trait space circular caused the time averaged fitness to strictly decrease with exploration rate, indicating that in the absence of asymmetry, there is no benefit to an intermediate level of exploration. Multidimensional trait spaces could also be considered, but without dependencies between the axes, this may reduce to several copies of a one dimensional trait. The fitness functions could also be changed within this framework. For example, one could consider fitness that comes from nonlinear or multiplayer games, like the Public Goods Game, or stochastic fitness functions, such as jumping to a random position at some constant frequency, or some constant positions with some random frequency \cite{carja2014evolution, matic2019mutation}. One could even make the exploration rate itself non-constant, possibly modeling it as a decreasing function of time, such as a linear or exponential function, and study the evolution of the parameters of these functions. Lastly, one could experiment with more intelligent agents. Whereas agents in this model explored randomly, one could use a reinforcement learning framework like Q-learning to model agents who explore based on previous knowledge. Such modifications would certainly change the balance of importance between exploration and exploitation, likely leading to different results. 


\section*{Acknowledgements}
B.M. is supported by a Dartmouth Fellowship. F.F. is supported by the Bill \& Melinda Gates Foundation (award no. OPP1217336), the NIH COBRE Program (grant no. 1P20GM130454), a Neukom CompX Faculty Grant, the Dartmouth Faculty Startup Fund, and the Walter \& Constance Burke Research Initiation Award. 

\section*{Competeing Interests}
The authors have no competing interests to declare. 

\newpage

\printbibliography

\end{document}